\documentclass[aps,prl,reprint,groupedaddress,twocolumn,showpacs,amsmath,superscriptaddress]{revtex4-1}
\usepackage{amssymb}
\usepackage{graphicx}
\usepackage{bm}
\usepackage{hyperref} 

\def\mynote#1{#1}
\def\mychange#1{#1}

\begin{document}

\title{Nonparaxial abruptly autofocusing beams}
\author{Raluca-Sorina Penciu}
\affiliation{Department of Mathematics and Applied Mathematics, University of Crete, 70013 Heraklion, Crete, Greece}
\author{Konstantinos G. Makris}
\affiliation{Crete Center for Quantum Complexity and Nanotechnology, Department of Physics, University of Crete, 71003, Heraklion, Greece.}
\author{Nikolaos K. Efremidis}
\email[Corresponding author: ]{nefrem@uoc.gr}
\affiliation{Department of Mathematics and Applied Mathematics, University of Crete, 70013 Heraklion, Crete, Greece}

\date{\today}


\begin{abstract}
We study nonparaxial autofocusing beams with pre-engineered trajectories. We consider the case of linearly polarized electric optical beams and examine their focusing properties such as contrast, beam width, and numerical aperture. Such beams are associated with larger intensity contrasts, can focus at smaller distances, and have smaller spot sizes as compared to the paraxial regime. 
\end{abstract}

\maketitle 

Since 2007 when exponentially truncated diffraction-free Airy beams have been predicted and observed~\cite{sivil-ol2007,sivil-prl2007} the study of curved and accelerating beams has attracted a lot of attention. Due to their unique properties such beams are associated with a variety of potential applications in areas including imaging~\cite{jia-np2014,vette-nm2014}, filamentation~\cite{polyn-science2009,polyn-prl2009}, particle manipulation~\cite{baumg-np2008,zhang-ol2011}, and plasmon generation~\cite{salar-ol2010} (see also the review~\cite{hu-springer2012}). 

Abruptly autofocusing (AAF) waves are accelerating waves that suddently generate a focal spot, in the sense that the intensity of light remains almost constant during propagation up until the focal point where it abruptly increases by several orders of magnitude~\cite{efrem-ol2010}. The principle of operation relies on the radial caustic collapse at the focal spot. Such beam were observed in~\cite{papaz-ol2011,zhang-ol2011} and utilized for creating ablation spots~\cite{papaz-ol2011} and in particle manipulation~\cite{zhang-ol2011}. Originally AAF beams relied on radial Airy waves, but subsequently different classes following pre-engineered caustic trajectories were suggested~\cite{chrem-ol2011}. The generation of AAF waves in the Fourier space is much simpler requiring the application of a phase mask to a slowly varying amplitude~\cite{chrem-ol2011-fourier,chrem-pra2012}. Utilizing the same principles autodefocusing beams (whose intensity drops by orders of magnitude right after a hot spot) and bottle beams \mynote{(high intensity closed light surfaces)} have been suggested~\cite{chrem-pra2012}. AAF waves have been utilized for the controlled filament generation at particular spatial locations~\cite{panag-nc2013}. By inducing a vortex phase to the autofocusing beams it is shown that the focus takes the form of a vortex ring~\cite{davis-oe2012,jiang-oe2012}. The polarization degree of freedom in generating autofocusing beams has been explored in~\cite{liu-ol2013,wang-apb2014}.  

In the nonparaxial regime accelerating \mynote{waves} offer the possibility of bending at large angles~\cite{froeh-oe2011,kamin-prl2012}. Such accelerating beams with circular profiles have the form of a filtered Bessel function~\cite{kamin-prl2012,courv-ol2012,zhang-ol2012}. Accelerating beams with parabolic and elliptic trajectories can be expressed in terms of Mathieu and Weber functions~\cite{zhang-prl2012,aleah-prl2012,bandr-njp2013}. In~\cite{penci-ol2015} closed form expressions for the input phase required for nonparaxial accelerating beams with different trajectories were found. 

In this work we utilize the properties of nonparaxial accelerating beams in abruptly autofocusing and autodefocusing waves. We consider the case of linearly polarized (LP) electric modes that follow different trajectories. We systematically analyze the focusing properties of these beams, such as the amplitude of the field components, the beam full width at half maximum (FWHM), the aperture size, and the amplitude of the initial profile. We find that nonparaxial AAF waves can have larger intensity contrasts, can focus at smaller distances, and have smaller spot sizes as compared to the paraxial regime. The presence of an optical field component in the propagation direction results is a focal spot with elliptic profile. 

We consider the beam propagation in a homogeneous dielectric medium under linear conditions. Utilizing Gauss' law $\nabla \cdot {\bm D} =0 $, the electric field is expressed as
\begin{align}
{\bm E}=-\frac{1}{\epsilon} \nabla\times {\bm F} , \label{eq:E0}
\end{align}
where ${\bm F}$ is an auxiliary vector potential and $\epsilon$ the electric permittivity. In the case of monochromatic waves the vector potential satisfies the Helmholtz equation 
\begin{equation}
\nabla^2 {\bm F} + k^2 {\bm F} = 0 \label{Helmholtz},
\end{equation}
where $k=n\omega/c=2 \pi/\lambda$, $\nabla^2=\partial_x^2+\partial_y^2+\partial_z^2$, \mynote{$(x,y)$ and $(r,\theta)$ are the transverse coordinates in Cartesian and polar form}, and $z$ is the propagation coordinate. We assume that at the input plane ($z=0$) the $\bm F$ potential if polarized along the $y$ direction and is radially symmetric. Due to the symmetries of the Helmholtz equation the auxiliary vector potential maintains its radial profile upon propagation, i.e., $\bm F = \hat{\bm y}F(r,z)$. Substitution to Eq.~(\ref{eq:E0}) then leads to 
\begin{align}
{\bm E}=(1/\epsilon)(\hat{{\bm x}} \partial_z F
-\hat{{\bm z}} \cos\theta \partial_r F).
\label{eq:E}
\end{align}
In Eq.~(\ref{eq:E}) we notice that the $x$ component of the electric field is radially symmetric, whereas the $z$ component exhibits a dipolar structure. Thus, in the rest of the paper we prefer to depict the intensity along the $x-z$ plane where the amplitude attains a maximum as a function of $\theta$. 

Within the ray optics approximation the solution of Eq.~(\ref{Helmholtz}) takes the form $\bm F(\bm r)=A(\delta\bm r)e^{i(\bm k\cdot(\bm r-\bm\rho)+\phi(\bm\rho)+g(\delta(\bm r-\bm\rho)))}$ where $\bm\rho=(\xi,\eta,0)$ is a point at the initial plane and $\delta$ is a small parameter. 
Then from Eq.~(\ref{eq:E0}) we obtain an approximate expression for the optical field
\begin{equation}
\bm E\approx (i/\epsilon)[\hat{\bm x}k_zF-\hat{\bm z}k_xF]. 
\label{eq:Eapprox}
\end{equation}
In the $x-z$ plane $k_y$ is zero and thus the intensity 
\begin{equation}
I(x,y=0,z)=|\bm E|^2\approx k^2|F(x,y=0,z)|^2/\epsilon^2
\label{eq:intensity}
\end{equation}
is proportional to $|F|^2$. 
This expression is highly accurate except from regions where the amplitude features are of the order of the wavelength. 
Furthermore, since along the $x-z$ plane $k_x=\phi_\rho(\rho)=\phi_\xi(\xi)$ and $k_z=\sqrt{k^2-k_x^2}$ which can be utilized to obtain approximate expressions for the electric field components by utilizing the exact formulas of~\cite{penci-ol2015}. \mynote{In the rest of the paper we prefer to numerically compute Eq.~(\ref{eq:E}) although the comparison with the exact expressions given by Eqs~(\ref{eq:Eapprox})-(\ref{eq:intensity}) is excellent.} 
\mynote{Our numerical algorithm consists of exactly solving Helmholtz Eq.~(\ref{Helmholtz}) in the Fourier space using Hankel transforms. Then the electric field is directly compute using Eqs.~(\ref{eq:E}).}

\begin{figure}
\centering
\includegraphics[width=\columnwidth]{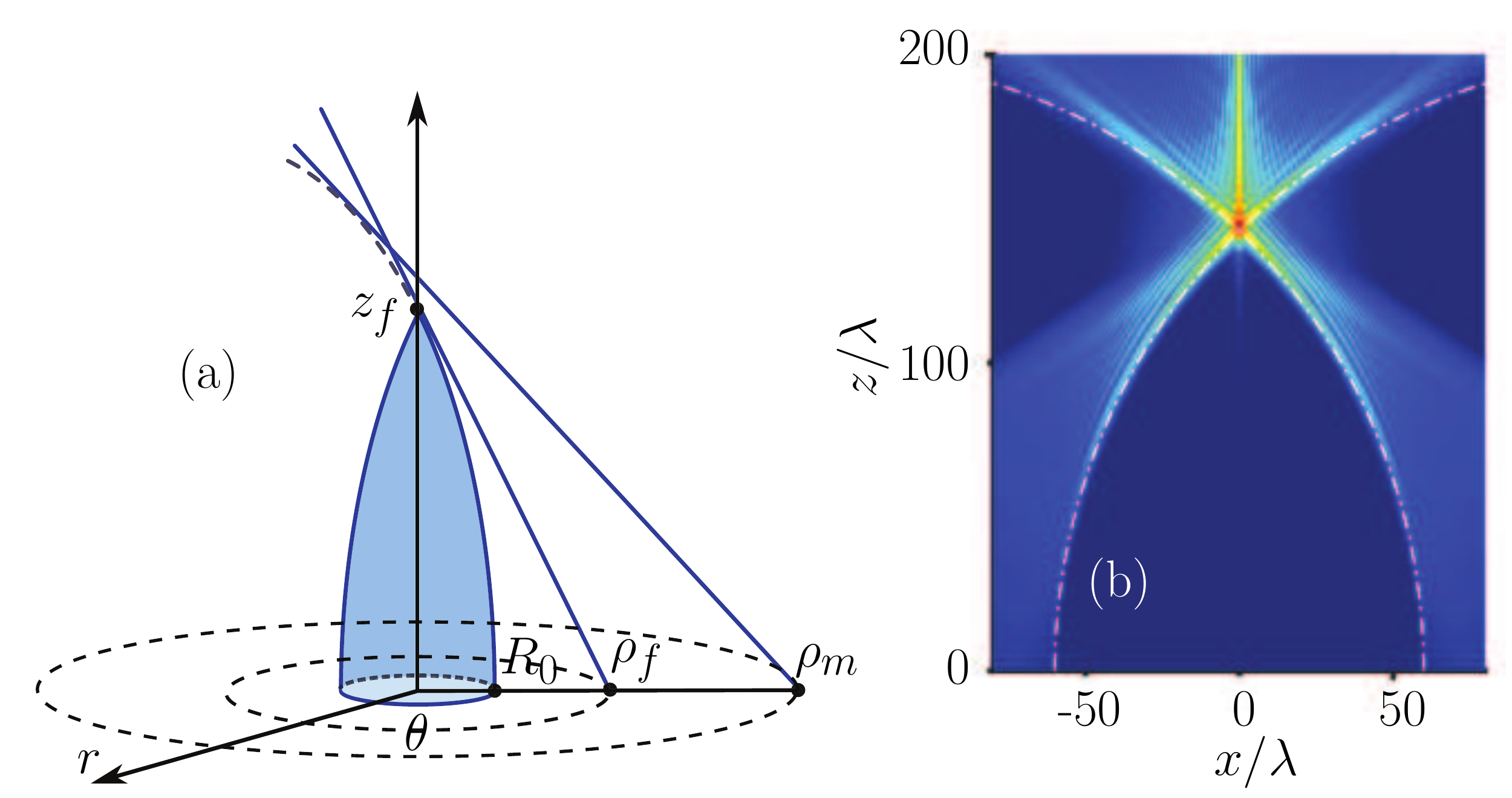} 
 \caption{(a) Ray-optics schematics of a nonparaxial autofocusing beam. The initial wave extends from $R_0$ to $\rho_m$. The conical bundle of rays starting from $\rho_f$ intersects at the focus $z_f$. (b) Intensity evolution (in logarithmic scale) in the $(x,z)$ plane for a circular autofocusing beam \mynote{(the parameters are the same as those in Fig~\ref{Fig1_circ})}. The white dot-dashed lines represent the caustic surface. \label{fig:1}}
\end{figure}
The geometry of the nonparaxial autofocusing beams is shown in Fig~\ref{fig:1}(a). At the input plane the beam extends between $R_0<r<\rho_m$; The inner part of the initial condition is a void disc with radius $R_0$ and the aperture is defined by $\rho_m$. The rays launched from an arbitrary radius at the input plane $\rho$ are tangent to the caustic surface at $(r_c(\rho),z_c(\rho))$. Such rays tangent at the focus $z_f$ are launched from a circle with radius $\rho_f$ (thus $\rho_m\ge\rho_f$) leading to an abrupt increase of the intensity profile at the focus [see Fig.~\ref{fig:1}(b)]. 

\begin{figure}[t]
\centering
\includegraphics[width=\columnwidth]{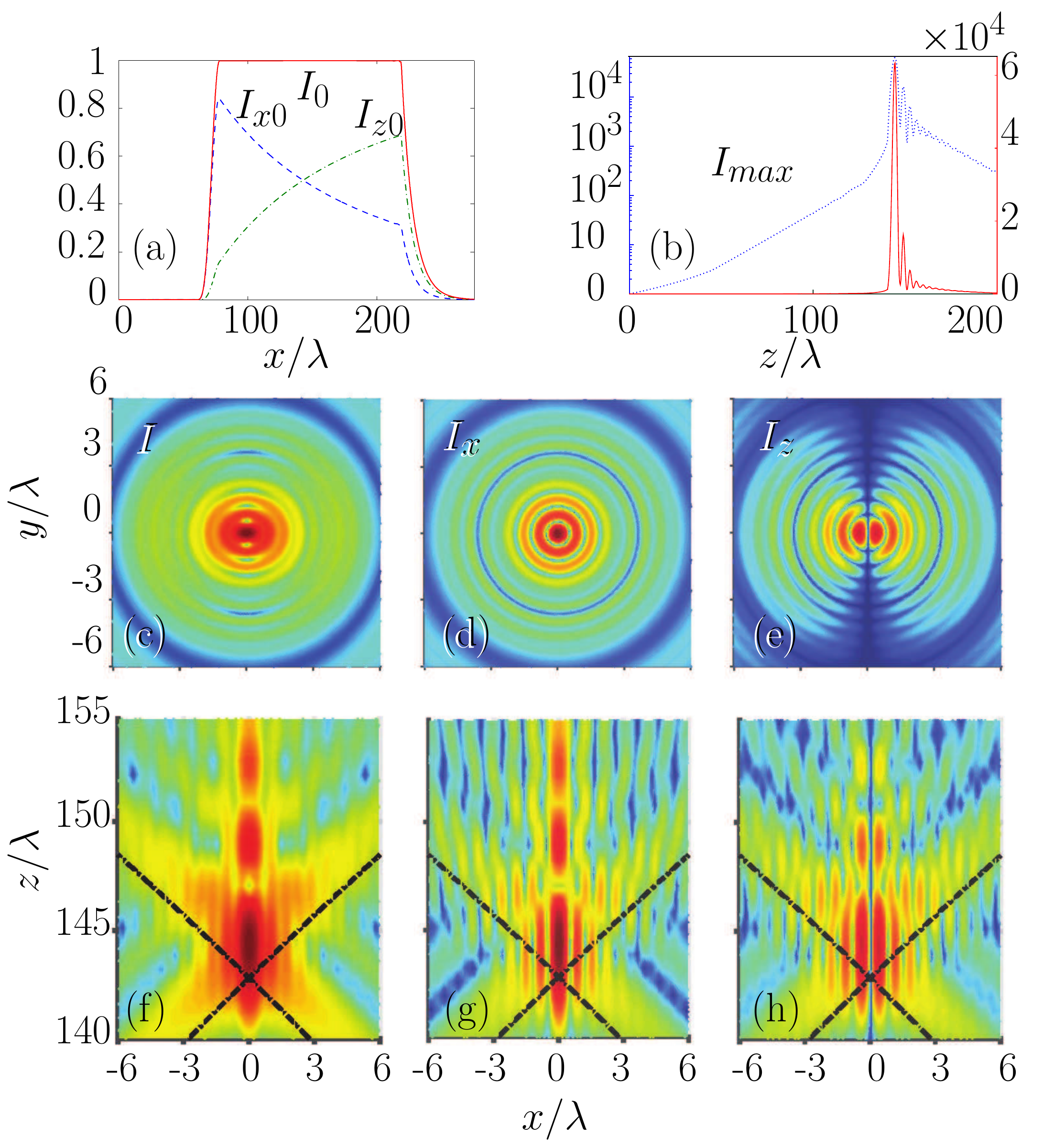} 
 \caption{AAF beams following a circular trajectory with $R=200 \lambda$ and $R_0 = 60 \lambda$. (a) Initial intensity in the $x$ direction: $I_0$ (red solid curve), $I_{x0}$ (blue dashed curve) and $I_{z0}$ (green dotted dashed curve). (b) Evolution of the maximum intensity contrast in linear (red solid) and logarithmic (blue dotted) scales. Intensity distribution at the focal plane: (c) $I$, (d) $I_x$, (e) $I_z$ (logarithmic scale). Intensity evolution in the $(x,z)$ plane near the focus: (f) $I$, (g) $I_x$, (h) $I_z$ (logarithmic scale). The black dotted dashed lines are the predicted circular caustics. \label{Fig1_circ} }
\end{figure}

We can engineer autofocusing beams with desired trajectories and intensity contrasts at the focus by selecting the amplitude and the phase of the vector potential at the input plane as $F(\rho)=A(\rho)e^{i\phi(\rho)}$. 
\mynote{We define the intensity contrast as the maximum intensity at the transverse plane ($z=\mathrm{constant}$) divided by the maximum intensity at the input plane $C(z)=I_{\mathrm{max}}(z)/I_{\mathrm{max}}(0)$. In all our simulations we set $I_{\mathrm{max}}(0)=1$ and thus $C(z)=I_{\mathrm{max}}(z)$}.  
\mynote{Since the rays cannot distinguish between Cartesian $(x,z)$ and radially symmetric cylindrical coordinates $(r,z)$,} the expressions for the phase are exactly the same as in the 1+1D case. Thus we can directly adopt the results of~\cite{penci-ol2015} where analytic expressions for the phase were provided to generate nonparaxial caustics with different trajectories. 
\mynote{Note that the utilization of ray optics in our analysis implies that the scale of the trajectories is larger than the wavelength.}
Then we independently select the amplitude of the autofocusing beam. 
Unless stated otherwise, the amplitude of the initial vector potential is zero for $\rho < R_0 $ or $\rho > \rho_m$ while it attains smoothly a constant value inside the interval $R_0 < \rho < \rho_m$, as shown for example in the solid line of Fig.~\ref{Fig1_circ}(a) [due to Eq.~(\ref{eq:intensity}) along the $x$ direction $|F|^2$ and $|E|^2$ are almost proportional]. 

We first study nonparaxial autofocusing light beams, following the circular trajectory $r=\sqrt{R^2-z^2}-(R-R_0)$ with radius $R$. The value of $\rho_m$ is selected to be large enough, so that it does not significantly affect the intensity contrast 
(we are going to separately study the effect of $\rho_m$ in more detail later on). 
From Eq.~(\ref{eq:Eapprox}) we find that \mynote{$I_x=|E_x|^2\approx k_z^2|F|^2/\epsilon^2$} and \mynote{$I_z=|E_z|^2\approx k_x^2|F|^2/\epsilon^2$} where $k_x=k_\rho\cos\theta$, $k_z=(k^2-k_\rho^2)^{1/2}$, and  $k_\rho=k[(\rho+R-R_0)^2-R^2]^{1/2}/(\rho+R-R_0)$. The field component amplitude behavior becomes apparent by noting that at the input plane as $\xi$ increases the rays bend at larger angles and thus the $x$ and $z$ field components decrease and increase, respectively. 

In Fig.~\ref{Fig1_circ}(b) we show the evolution of the maximum intensity along the $z$ direction, with an intensity contrast of about $6.5 \times 10^4$ at the focus.
We note that the intensity remains \mynote{at low levels} up until the focal point. Then, \mynote{even in the logarithmic scale,} a knee appears and abruptly the intensity increases by several orders of magnitude. In Figs.~\ref{Fig1_circ}(c)-(e), we depict the intensity profile at the focal plane ($z=z_f$). 
\mychange{Note that $E_x$ is radially symmetric and exhibits its maximum on axis, while $|E_z|$ is zero on the $y$-axis and its dipolar structure is characterized by two maxima centered symmetrically with respect to the $y$ axis.}
Thus, the resulting total beam intensity is elliptic with FWHM diameters $(0.81, 0.52)\lambda$ in the transverse directions. In Figs.~\ref{Fig1_circ}(f)-(h) we present the field intensity in the $(x,z)$ plane in the vicinity of the focal point.

\begin{figure}[t]
\centering
\includegraphics[width=\columnwidth]{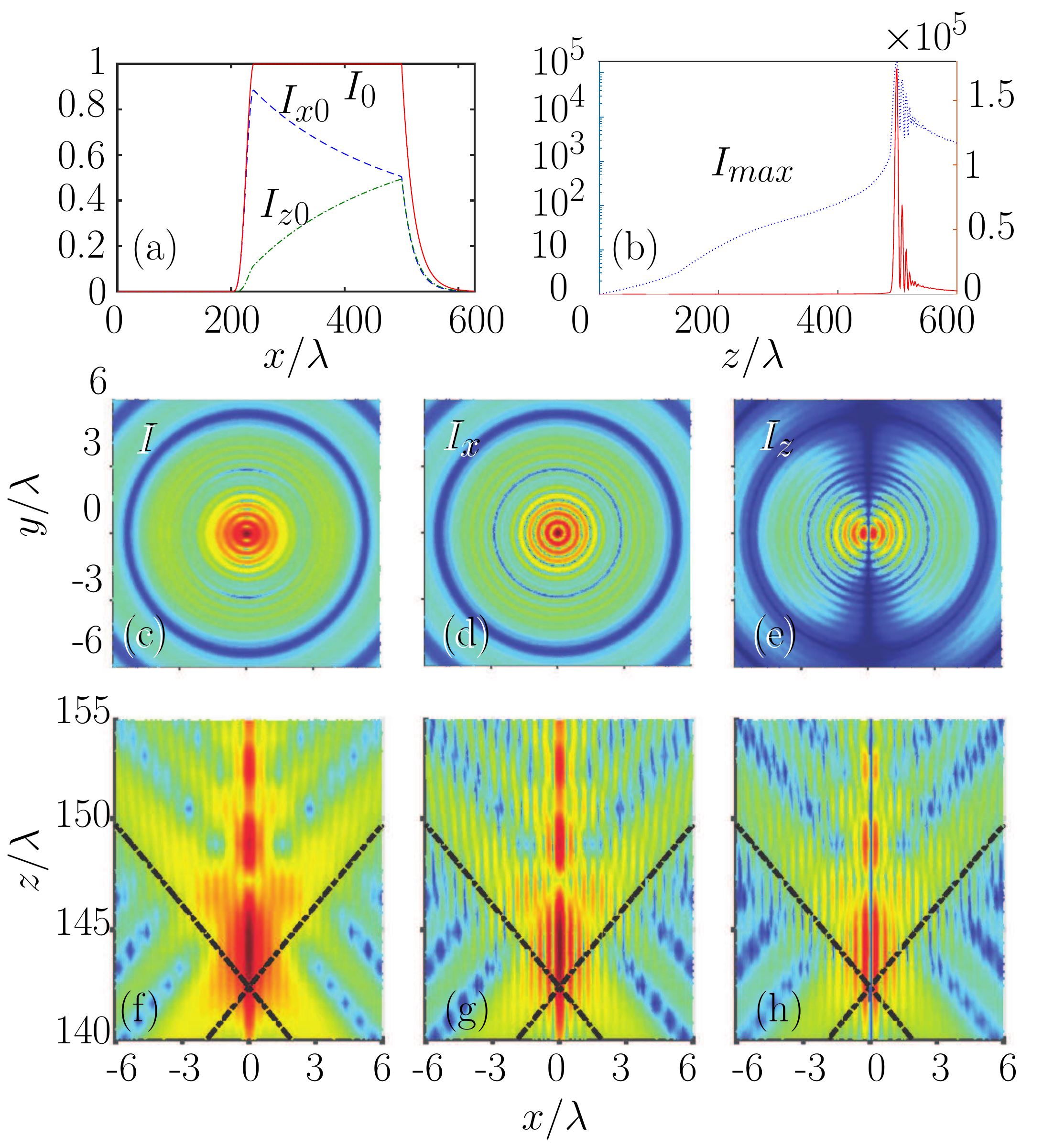} 
\caption{Same as in Fig.~\ref{Fig1_circ} for an AAF beam following a parabolic trajectory with $R_0=200 \lambda$ and $z_f=500 \lambda$. \label{Fig6_power} }
\end{figure}
In Fig.~\ref{Fig6_power} we show results similar with those shown in Fig.~\ref{Fig1_circ} this time for a parabolic autofocusing beam. The field distribution has the same characteristics as in Fig.~\ref{Fig1_circ}, while the intensity contrast at the focal point is increased up to $1.7 \times 10^5$. In the transverse plane the FWHM is $(0.74,0.57)\lambda$. The main difference of such power law caustics is that ideally the rays never bend up to $90^\circ$. Thus for the same trajectory can extend to large distances.

\begin{figure}
\centering
\includegraphics[width=\columnwidth]{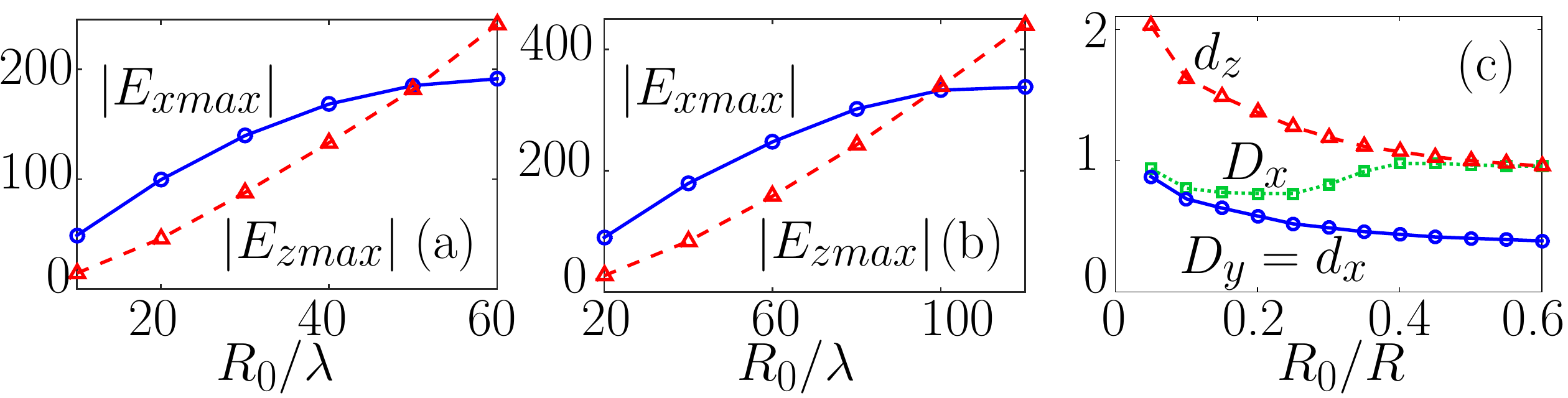} 
 \caption{(a)-(b) Maximum amplitude of the field components at the focus [$|E_{x max}|$ (blue solid line with circles) and $|E_{z max}|$ (red dashed line with triangles)] normalized to the initial maximum absolute value of the total field for circular autofocusing beams with (a) $R=100 \lambda$  and (b) $R=200 \lambda$ vs the initial radius of the beam $R_0/\lambda$. (c) FWHM of the field components $(d_x,d_z)$ and the total field $(D_x,D_y)$.\label{circ_Exmax_Ezmax} }
\end{figure}
It is highly desirable to generate a focal spot that is as small as possible with the highest contrast. In Fig~\ref{circ_Exmax_Ezmax}(a)-(b) we see the maximum amplitude of AAF beams following a circular trajectory as a function of $R_0$. By increasing $R_0$ the amplitude of both components increases. For smaller values of $R_0$ the $x$ component is stronger than the $z$ component. However, there is a point where, due to increased bending of the rays, the $z$ component becomes stronger than the $x$ component. 
The FWHM depicted in Fig.~\ref{circ_Exmax_Ezmax}(c) are the same for both cases shown in Figs.~\ref{circ_Exmax_Ezmax}(a)-(b). Thus, the beam diameters are mainly angle dependent (rather than scale dependent) variables at least for beam dimensions much larger than the wavelength. 
A strong $z$ component increases the ellipticity of the focal spot as shown in Fig.~\ref{circ_Exmax_Ezmax}(c). The $x$ component has radial symmetry and thus its FWHM $d_x$ is the same along both transverse directions. The $z$ component of the field is zero along the $y$ direction. Along the $x$ direction its FWHM, $d_z$, takes into account both peaks of the dipole structure. 
The FWHM of the beam is then $D_y=d_x$ along the $y$ direction. 
Along the $x$ direction as $R_0$ goes to zero we reach the paraxial regime where the beam takes a radial profile and $D_x=d_x$. On the other hand, as $R_0/R$ increases the beam becomes elliptic elongated along the $x$ direction and $D_x$ approaches $d_z$. We note that the due to higher bending of the rays, along the $y$ direction the FWHM monotonically decreases with $R_0/R$ whereas the value of $D_x$ attains a minimum for an intermediate value of $R_0/R$.

\begin{figure}
\centering
\includegraphics[width=\columnwidth]{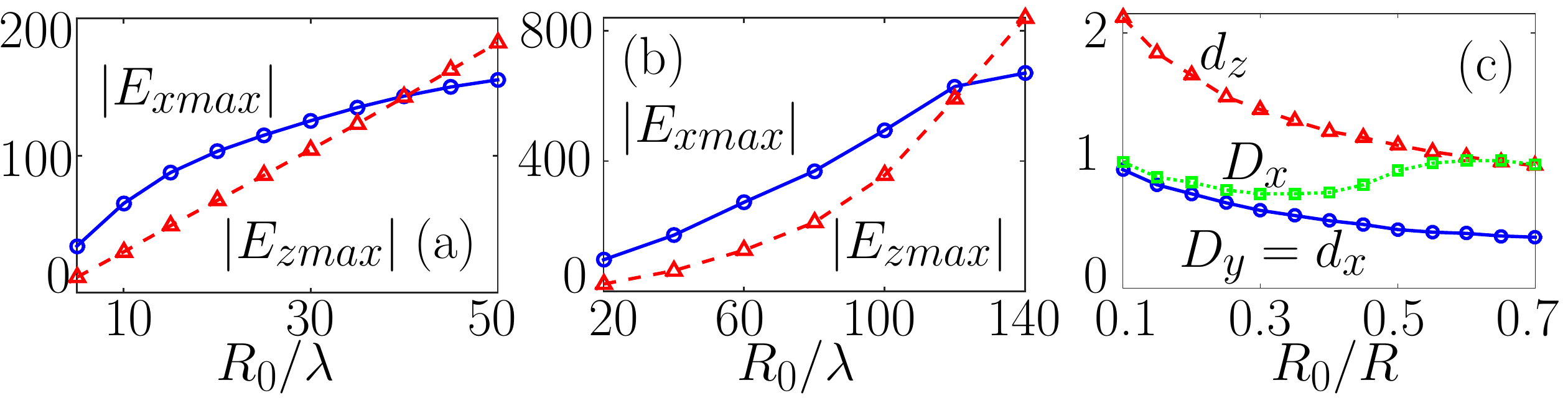} 
 \caption{Maximum field component contrasts at the focus [$|E_{x max}|$ (blue solid line with circles) and $|E_{z max}|$ (red dashed line with triangles)], normalized to the initial maximum absolute value of the total field, for elliptic autofocusing beams with radius $R=200 \lambda$ for (a) $\alpha=0.5$ and (b) $\alpha=1.5$ vs the initial radius of the beam $R_0/\lambda$.  (c) FWHM of the total field $(D_x,D_y)$  and its components $(d_x,d_z)$ for the case shown in (b).\label{ellip_Exmax_Ezmax} }
\end{figure}


The amplitude contrast of the field components as a function of the trajectory parameters is also depicted in Fig.~\ref{ellip_Exmax_Ezmax}(a)-(b) for elliptic autofocusing beams following the trajectory $r = \sqrt{R^2-(z/\alpha)^2} - (R - R_0)$, with $R=200 \lambda$ for $\alpha = 0.5$ (major axis along the $x$ direction) and $\alpha =1.5$ (major axis along the $z$ direction). We see that when the major axis of the ellipse is in the propagation direction, $R_0$ can reach higher values before $|E_{z max}|$ surpasses $|E_{x max}|$ and the electric field contrast is more pronounced. In Fig.~\ref{ellip_Exmax_Ezmax}(c) the beam diameters of the field and its components is shown for $\alpha=1.5$. 


\begin{figure}
\centering
\includegraphics[width=\columnwidth]{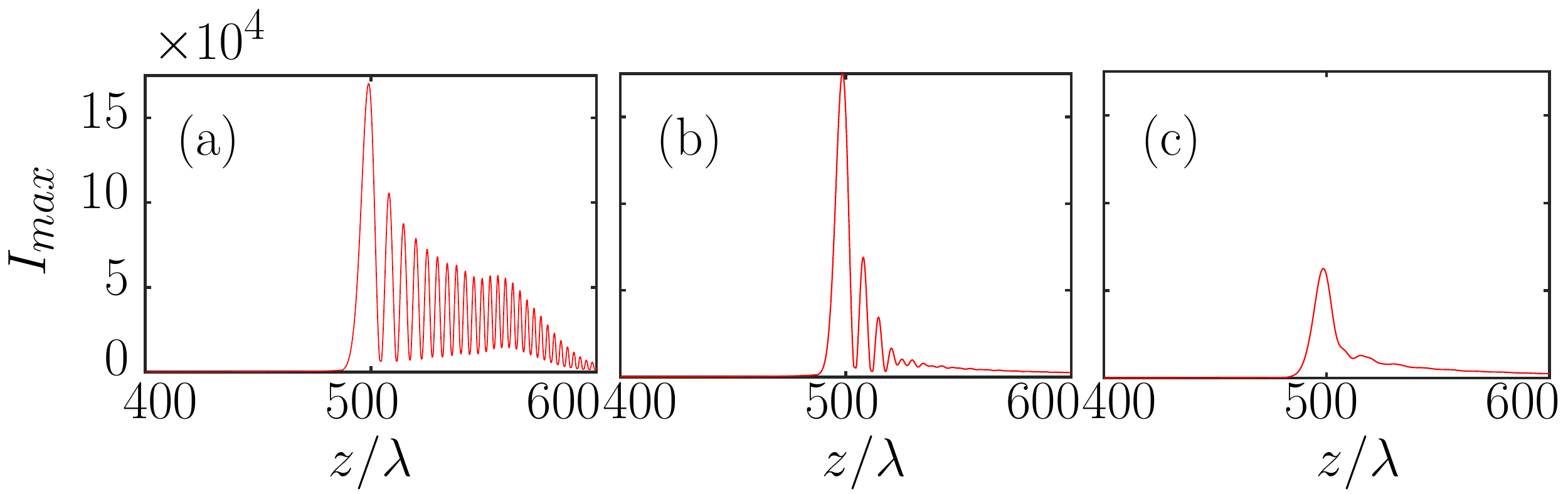} 
 \caption{Maximum intensity contrast as a function of the propagation distance for a parabolic autofocusing beam with initial radius $R_0=200 \lambda$ and focusing distance $z_f=500 \lambda$, with numerical aperture (a) $\rho_m=2 \rho_f$,  (b) $\rho_m=5/4 \rho_f$ and (c) $\rho_m= \rho_f$.\label{2xf_15xf}}
\end{figure} 
The decrease in the maximum intensity after the focus is not monotonic, but it exhibits oscillations leading to a slower decay of the autofocusing beam, as can be seen in Fig.~\ref{2xf_15xf}(a) in the case of a parabolic trajectory. It would be desirable to be able to reduce the intensity of the laser beam after the focus as fast as possible. Noting that the rays contributing to such oscillations are generated after $\rho_f$, a simplified solution (that does not take into account diffraction) would be to set the amplitude to zero for $\rho>\rho_f$. As shown in Fig.~\ref{2xf_15xf}(c) the elimination of the ray after $\rho_f$ leads to increased diffraction of the part of the beam that contributes to the focus and thus to the reduction of the intensity contrast. An optimum value of the aperture is one that reduces as much as possible the intensity oscillations after the focus, without significantly reducing the value of the intensity contrast. Specifically, we see in Figs.~\ref{2xf_15xf}(a)-(b) that for an aperture $\rho_m=2\rho_f$ and $\rho_m=5\rho_f/4$ the intensity contrast remains almost the same. However, for $\rho_m=5\rho_f/4$ the intensity oscillations after the focus are significantly reduced. 

\begin{figure}
\centering
\includegraphics[width=\columnwidth ]{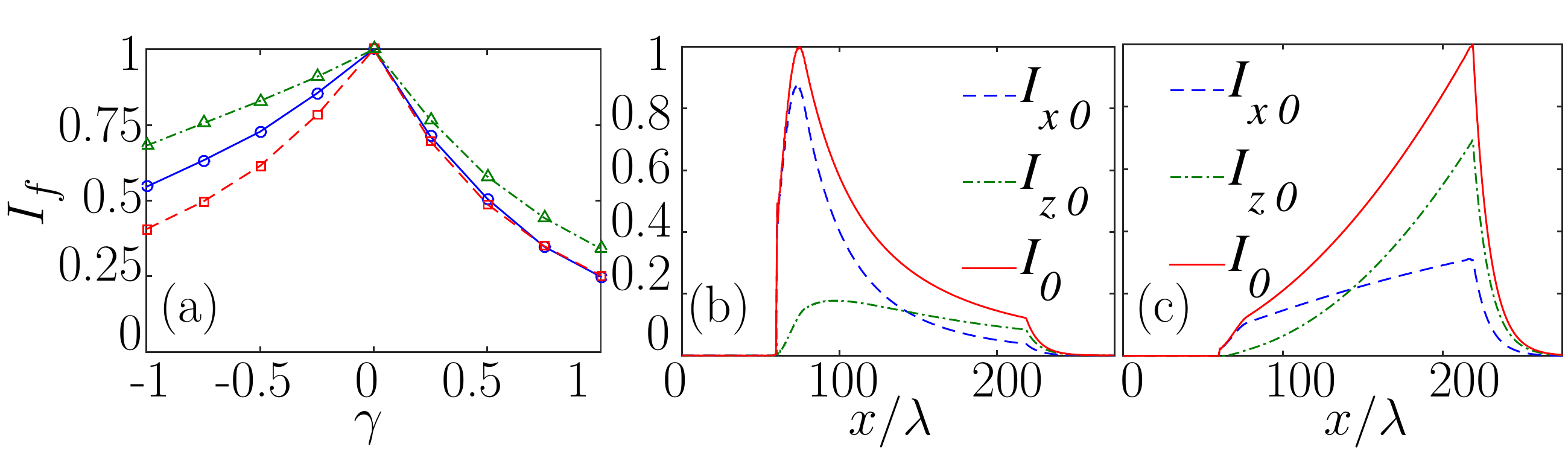} 
\caption{(a) Intensity contrast at the focus normalized to its value at $\gamma=0$ vs $\gamma$ for a circular autofocusing beam with $R=200 \lambda$, $R_0=60\lambda$ (blue solid line with circles), elliptic autofocusing beam with $\alpha=0.5$, $R=200\lambda$, $R_0=30\lambda$ (red dashed line with squares) and parabolic autofocusing beams with $R_0=200\lambda$, $z_f=500 \lambda$ (green dot-dashed line with triangles). Initial field intensity for the circular autofocusing beam with $R=200\lambda$ ,$R_0=60\lambda$ for the initial vector potential amplitude with (b) $\gamma=1$ and (c) $\gamma=-1$, respectively.\label{gamma}} 
\end{figure}
We now explore the influence of the initial amplitude of the vector potential on the intensity contrast. We assume an amplitude of the form $A(\rho) = \rho^{-\gamma}$ for $R_0\leq \rho \leq \rho_m$ that smoothly going to zero elsewhere. By varying the amplitude exponent $\gamma$ as shown in Fig.~\ref{gamma} we find that the maximum intensity contrast at the focus is reached for $\gamma = 0$ for different classes of trajectories. 
The contrast is enhanced for $\gamma=0$ because for this value the amplitude is maximized in the initial plane at $\rho_f$. 

\begin{figure}
\centering
\includegraphics[width=0.9\columnwidth]{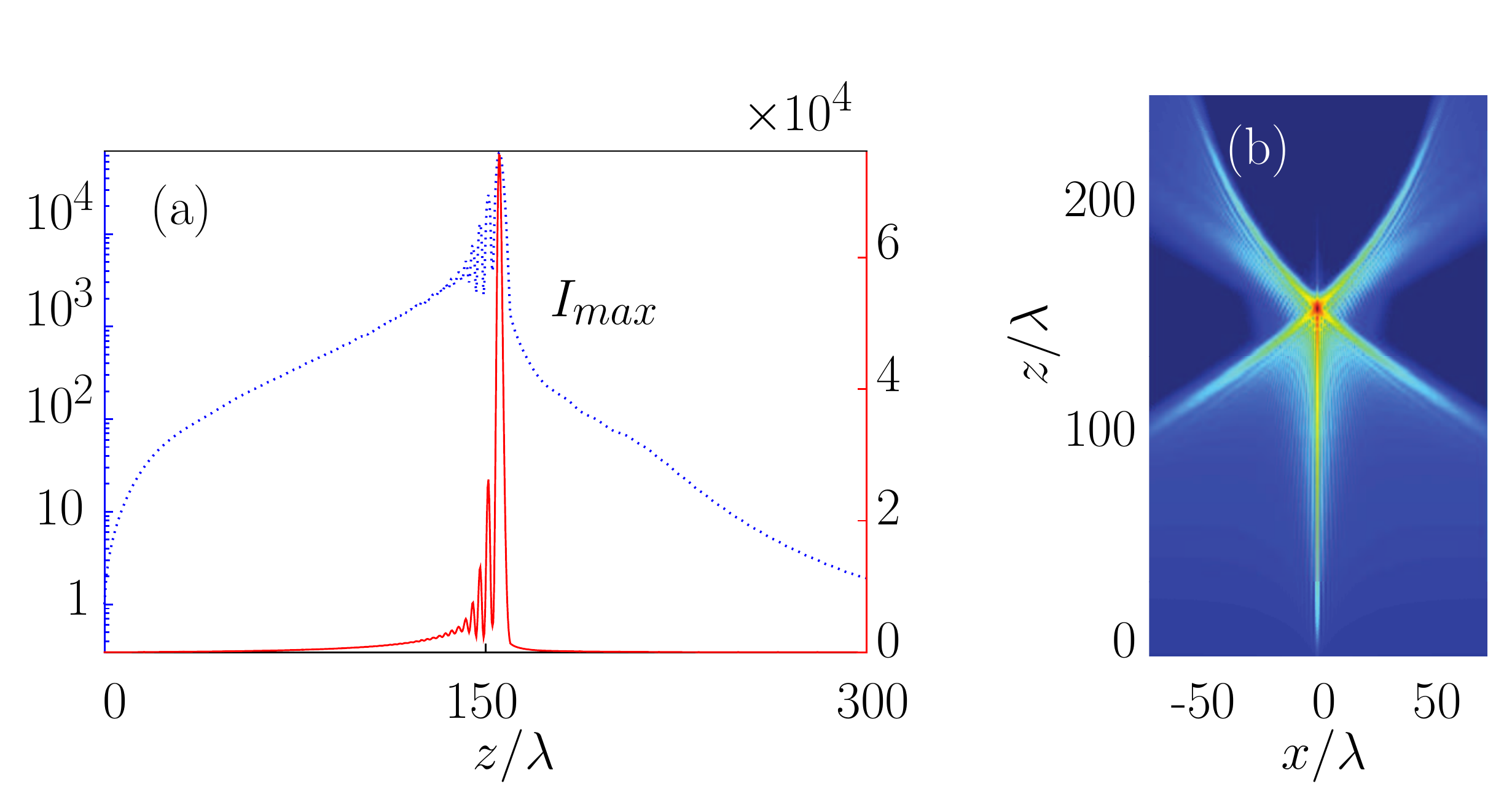} 
\caption{An abruptly autodefocusing beam with $R=200\lambda$, $R_0=60\lambda$, and $z_0=300\lambda$. (a) Maximum intensity contrast vs propagation distance and (b) intensity dynamics (in logarithmic scale)\label{fig:defocusing}}
\end{figure}
The same principles described before can be applied for the generation of abruptly autodefocusing beams. The intensity profile of such beams decreases rapidly after the focus by several orders of magnitude. Abruptly autodefocusing beams can be utilized in burning the surface of a material without affecting its bulk. A typical example is shown in Fig.~\ref{fig:defocusing} for a circular trajectory 
$r = f(z) = \sqrt{R^2- (z_0-z)^2}-(R-R_0)$.

In conclusion, we have studied nonparaxial autofocusing and autodefocusing beams with pre-engineered trajectories. For linearly polarized electric fields, we have studied their focusing properties such as contrast, beam width, and numerical aperture. We have found that such beams are associated with larger intensity contrasts, focusing at smaller distances, and smaller spot sizes as compared to the paraxial case. 

Supported by the Research Project ANEMOS co-financed by the European Union (European Social Fund-ESF) and Greek national funds through the Operational Program ``Education and Lifelong Learning'' of the National Strategic Reference Framework (NSRF)-Research Funding Program: Thales. 
The paper and the participation of N.K.E. has been made in the framework of the ``Erasmus Mundus NANOPHI project, contract number 2013-5659/002-001''. 
K.G.M. is supported by the European Union Seventh Framework Programme (FP7-REGPOT-2012-2013-1) under grant agreement 316165.


\newcommand{\noopsort[1]}{} \newcommand{\singleletter}[1]{#1}

\newpage

\end{document}